\documentclass[%
reprint,
superscriptaddress,
amsmath,amssymb,
aps
]{revtex4-2}

\usepackage[colorlinks=true,linkcolor=red,citecolor=blue,urlcolor=blue]{hyperref}
\usepackage{graphicx}
\usepackage{dcolumn}
\usepackage{stix}
\graphicspath{{./Figures/}}
\usepackage{amsmath,amssymb,physics}
\usepackage{cleveref}
\usepackage{multirow}
\usepackage{array}
\usepackage{mathtools}
\usepackage{orcidlink}
\usepackage{lineno}

\newcommand{\BFF}[1]{{\color{black}{{#1}}}}

\begin{document}
	
	\preprint{APS/123-QED}
	
	\title{Prospect of unraveling the first-order phase transition in neutron stars with $f$ and $p_1$ modes} 
	
	\author{Pratik Thakur \orcidlink{0000-0001-5902-7695}}
	\email{thakur.16@iitj.ac.in}
	\affiliation{Indian Institute of Technology Jodhpur, Jodhpur 342037, India}
	\affiliation{Indian Institute of Science Education and Research Bhopal, Bhopal 462066, India}
	
	\author{Sagnik Chatterjee \orcidlink{0000-0001-6367-7017} }
	\email{sagnik18@iiserb.ac.in}
	\affiliation{Indian Institute of Science Education and Research Bhopal, Bhopal 462066, India}
	
	\author{Kamal Krishna Nath \orcidlink{0000-0002-4657-8794}}
	\email{kknath@niser.ac.in}
	\affiliation{School of Physical Sciences, National Institute of Science Education and Research, An OCC of Homi Bhabha National Institute, Jatni 752050, India}
	
	\author{Ritam Mallick \orcidlink{0000-0003-2943-6388}}
	\email{mallick@iiserb.ac.in}
	\affiliation{Indian Institute of Science Education and Research Bhopal, Bhopal 462066, India}

	\date{\today}
	
	\begin{abstract}
		Quasi-normal modes of neutron stars are an exciting prospect for analyzing the internal composition of NSs and studying matter at high densities. In this work, we focus on studying the $f$- and $p$- quadrupolar oscillation modes, which couple with gravitational waves. We construct two different equation of state ensembles, one without and one with a first-order phase transition, and examine how $f$- and $p$-modes might help us differentiate them. We find ensemble specific exclusion regions in the $65\%$ and $95\%$ confidence contours of the frequency-damping time relations. The exclusion regions become more prominent for the higher-order oscillation modes. However, these modes have higher frequencies, which are beyond the detection capabilities of present gravitational wave detectors. The quasi-universal relations of dimensionless quantities prove to be ineffective in differentiating the equation of state ensembles, as they obscure the details of the equation of state.
		
	\end{abstract}
	
	\maketitle
	\section{Introduction}\label{sec:intro}
	Gravity compresses matter inside neutron stars (NSs) to densities several times larger than in atomic nuclei $n_s=0.16/{\rm fm}^3$, making them the densest objects in the current Universe.
	It is the strong nuclear force of Quantum Chromodynamics (QCD) that prevents a neutron star from gravitational collapse into a black hole.
	To this date, neutron stars and their mergers are the only available source of information for the behavior of matter under such extreme conditions, which remains inaccessible to terrestrial experiments and first principle calculations in QCD.
	
	A promising tool to access this information is to study neutron star oscillations about their stationary equilibria.
	There are various scenarios in which such oscillations can be excited, including core-collapse supernovae, inspiral and post-merger phases of binary neutron star mergers, close encounters of neutron stars with black holes and neutron star starquakes.
	Of particular interest are quadrupolar ($\ell=2$) modes since these can couple to the gravitational field and lead to the emission of gravitational waves (GWs), which current and next-generation GW observatories could observe. 
	
	There exists a plethora of different oscillation modes that can be classified by their dominant restoring force and their number of nodes inside the star.
	The phenomenologically most important one is the fundamental $f$-mode, \BFF{a non-radial ($\ell\geq2$)} breathing mode of the star with zero radial nodes.  
	Other examples are $g$- and $p$-modes, which are oscillation modes with arbitrary node numbers and for which gravity and gradients in the fluid pressure are the dominant restoring forces, respectively. \BFF{The g-modes arises due to the occurrence of a sharp discontinuity in the energy density due to a quark-hadron phase transition of the EOS \citep{Tonetto}. They may also arise in stars modelled with a temperature or composition gradient where the adiabatic speed of sound is not equal to the equilibrium speed of sound \citep{twoSounds_Jaikumar_2021}}. Finally, there are also w-modes, which are strongly damped GW modes that are dominated by variations of the spacetime metric \cite{kokkotasWModes1992}. The $p$- and $w$-modes have high frequencies ($5-12\ \rm kHz$) and are therefore probably not excited during neutron star mergers~\cite{Kokkotas:1999bd}.
	
	For stars with uniform density (with mass $M$ and radius $R$) in Newtonian gravity the $f$-mode can be computed in closed form $\omega_f^2=\frac{2\ell(\ell-1)}{2\ell+1}\frac{M}{R^3}$.
	The $f$-mode sits between $g$- and $p$-modes with frequencies $\nu_f=\omega_f/(2\pi)\approx1.3-2.8\ \rm kHz$ \cite{zhaoUniversalRelationsNeutron2022, Flores_2018} and damping times of $0.1-0.5\ \rm sec$ , and have a possibility of being detected with third generation GW detectors, the Einstein Telescope and the Cosmic Explorer \cite{Sathyaprakash_2019,Punturo_2010,Kalogera_2021}.
	
	The $f$-modes are known to correlate with the mean neutron star density $\omega_f\propto\sqrt{M/R^3}$~\cite{Andersson_towards_GW_astroseismo}.
	Furthermore, equation of state (EOS)-insensitive correlations between the dimensionless frequency $\Omega_f=GM\omega_f/c^3$, the dimensionless moment of inertia $\bar I=Ic^4/(G^2M^3)$ (G being the gravitational constant and c the speed of light) and the tidal deformability $\Lambda$ were found.
	The I-Love-Q relation~\cite{Yagi:2013awa, Nath_2023} implies a quasi-universal relation (UR) between $\Omega_f$ and $\Lambda$~\cite{chan_2014,Chirenti:2015dda,Lioutas_2021}. 
	Quasi-UR between quasi-normal mode (QNM) frequency and damping time correlates with the mass and radius of the star in EOS-independent manner \citep{Andersson_towards_GW_astroseismo} with significant improvements in recent works \citep{LKTsui_2005, Lioutas:2017xtn, Sotani_Kumar_2021} 
	using more robust and diverse EOS \citep{Benhar_2004,ranea-sandovalOscillationModesHybrid2018,Jaiswal_Chaterjee_2021, Aguirre_2022, Flores_2019, Pradhan_2021, Das_2021, kunjipurayilImpactEquationState2022,  Pradhan_2022}. The composition of NS (hadronic or quark) can give rise to bimodal UR \citep{Ranea_2019} with the UR getting affected by the rotation of the star \citep{Gaertig_PRL, Kruger_kokkotas_2020a,Kruger_kokkotas_2020b,Doneva_2013,Zink_2010,Yoshida_2012,Kastaun_2010, Passamonti, Pnigouras,Rosofsky,Breu:2016ufb,Musolino:2023edi} and also by the presence of magnetic fields \citep{Haskell:2013vha}.

	The novel ingredient in our work is to employ a large number of generic EOS models both smooth \citep{Annala:2019puf,Ecker_2022,chatterjee2023} and with a first-order phase transition (PT) \citep{gorda2023}, which are consistent with nuclear theory and perturbative QCD at low and high densities, respectively, with neutron star mass-radius measurements as well as with the bounds on the tidal deformability deduced from the GW170817 binary neutron star merger event.
	More specifically, we will compare two different EOS ensembles, one where the EOS is smooth and another which includes a first-order phase transition, which from here on we will refer to as ``smooth'' and ``with PT'', respectively. 
	Although there has been a number of work relating PT in neutron stars \citep{Bhattacharyya:2006vy, Chamel, veronica, Mallick:2018lbu, Orsaria_2019,Mallick:2020bdc,Kuzur:2021gnf,Prasad:2022dom, tews2, Reed,Haque:2022dsc,Pradhan:2023zmg}, the questions remains how one observationally separates stars which had undergone PT. \BFF{Flores and Lugones in 2014 \citep{Vásquez_Flores_2014} showed that hadronic and quark stars can indeed be differentiated using $p_1$ modes of oscillation. However, their results show that the frequency of oscillation for the hardonic and hybrid stars in both $f$- and $p_1$- lies in a similar range hence making the task of differentiating them very difficult. They further improved upon this and had extensively analyzed the $f$-modes bounds corresponding to extreme EOSs comprising a band \citep{Flores_2018}.}
	\BFF{Whereas, the properties of $g$-mode frequencies are dependent on the onset and width of the phase transition and an interesting work differentiating the fast and slow PT using $g$-mode has already been studied in the literature \citep{Tonetto}.}
	In this work we propose $f$- and $p$- modes as a tool to understand the presence or absence of first-order PT in EOSs and try to understand their implications on the quasi-URs. 
	
	The rest of the paper is structured as follows.
	In \cref{sec:EOS} we present the construction of our EOS models two ensembles.
	In \cref{sec:QNM} we discuss the neutron star QNM analysis.
	Then in \cref{sec:results} we present our results and finally summarize and conclude in \cref{sec:summary}.

	\section{Equation of State Modelling} \label{sec:EOS}
	In this work, we construct EOSs using an agnostic approach, for which we consider two cases. For the first one, we generate smooth EOSs using the randomization of the speed of sound and in the second case, we construct EOSs with a first-order PT.
	
	\subsection{EOS models without phase transition}

	We construct a family of EOSs by interpolating between chiral effective field theory (EFT) and perturbative QCD. The adiabatic speed of sound ($c_s$) is used as a parameter for interpolation at these intermediate densities. The $c_s$ also gives us the slope of the EOSs and helps to determine the EOS being bounded in the limit $0 < c_s \le 1$. At very low densities $n<0.5\,n_s$, we have used a tabulated version of the Baym-Pethick-Sutherland (BPS) model \citep{Baym1971}. Then, in the range $0.5\,n_s<n<1.1\,n_s$, we have constructed monotropes of the form $p(n)=K\,n^\Gamma$, where $K$ is fixed by matching to the BPS EOS. We sample $\Gamma$ uniformly $\in [1.77, 3.23]$ \citep{Altiparmak_2022} and ensure that the pressure remains entirely between the range defined by Hebeler et al. \citep{Hebeler_2013}. Between $1.1\,n_s < n \lesssim 40\,n_s$ we use the sound-speed parametrization method introduced in \citep{Annala:2019puf, Altiparmak_2022}.
	\BFF{In between $1.1n_s < n \le 40 n_s$ the number density is defined as  
		\begin{align}
			n(\mu) = n_1 \exp\left({\int_{\mu_1}^{\mu} \dfrac{d\mu'}{\mu' c_s^{2}(\mu')}}\right)
		\end{align}
		where $n_1 = 1.1 n_s$ and $\mu_1 = \mu(n_1)$
		The pressure can again be obtained from the number density as
		\begin{align}
			p(\mu) = p_1 + \int_{\mu_1}^{\mu} d\mu' n(\mu')
		\end{align}
		where the constant $p_1$ is the pressure at $n_1$
		To solve these two equations numerically we use a fixed number of segments between N(3,4,5,7) Ref\citep{Altiparmak_2022} and use a piecewise linear interpolation as:
		\begin{align}
			c_s^{2}(\mu) = \dfrac{(\mu_{i+1} - \mu) c_{s,i}^{2} + (\mu - \mu_i) c_{s,i+1}^{2}}{\mu_{i+1} - \mu_i}
		\end{align}
		where $\mu_i$ and $c_{s,i}^{2}$ being the chemical potential and the speed of sound sampled randomly between $\mu_1 \leq \mu_i \leq \mu_{N+1}$ and $0< c_{s,i}^{2} \leq 1$ at the $i$-th segment.}
	As the final step in our procedure, we keep solutions whose pressure, density, and sound speed at $\mu_{i}=2.6\,{\rm GeV}$ are consistent with the parametrized perturbative result for cold quark matter in beta-equilibrium \citep{Kurkela_2010}. More details on this prescription of EOS construction can be found in \citep{Altiparmak_2022}. 
	
	\BFF{
		The square of the speed of sound ($c_{s}^{2}$) for hadronic matter exhibits $\le 1/3$ at smaller densities and pQCD limits it to be $c_{s}^{2}=1/3$ where degrees of freedom are quarks. However, recent mass and radius measurement of various pulsars have led to the belief that the constraints of $1/3$ is violated at the cores of neutron stars \citep{Bedaque}. The peak in speed of sound usually suggest a smooth transition from hadronic matter to quark matter \cite{FERRER2023122608}. Reed and Horowitz \citep{Reed_cs} argued that hadronic matter can have $c_s^{2} \ge 1/3$ with some extreme exotic conditions causing stiffening of the EOS; however, it most likely indicate some other state than hadronic matter. 
		Tews et al. 2018 \citep{Tews_2018} argued that for stars to be composed entirely of hadronic matter it should have $c_s^{2} \le 1/3$. However, such EOSs will be unsuccessful in explaining higher mass stars, hence it is always expected that the speed of sound at intermediate densities is bound to cross the conformal limit. This would mean that the speed of sound parametrized EOSs can contain hadronic, quarkyonic or EOSs with smooth phase transition. The EOSs constructed in this manner will be referred to as `smooth' EOSs from here on.}

	\subsection{EOS models with first-order phase transition}
	
	The EOSs with first-order PT follow a similar construction till the chiral EFT band. From the chiral EFT band, we now chose two polytropes which will have a discontinuity in the energy density for the PT. Each of these polytropes has polytropic indices $\gamma_{1}$ and $\gamma_{2}$ which are chosen randomly from a uniform distribution. In the next step, we randomly chose the energy density at which the PT occurs ($\epsilon_{PT}$) and the thickness of the discontinuity $\Delta \epsilon$. The following method was used to construct the whole EOS.

	\begin{itemize}
		\item For density in the  range $\epsilon_{EFT} < \epsilon < \epsilon_{PT}$ we use a polytrope of the form $p_{1} (\epsilon) = \kappa_{1} \epsilon^{\gamma_1}$, where $\kappa_1$ is a constant.
		\item Next we have the discontinuity for first-order PT equivalent to $\Delta \epsilon$ where the pressure remains constant throughout.
		\item For $\epsilon_{PT}+\Delta \epsilon < \epsilon $ we use $p_2(\epsilon) = \kappa_2 \epsilon^{\gamma_2}$, where $\kappa_2$ is a constant
	\end{itemize} 
	
		We sample the values of $\gamma_1$ and $\gamma_2$ from a uniform distribution from [0,10]. The transition energy density ($\epsilon_{PT}$) is sampled in a range of [200,1000] $\text{MeV/fm}^{3}$ and the discontinuity in energy density ($\Delta \epsilon$) between [10,1100] $\text{MeV/fm}^{3}$. This process was repeated a number of times to generate the desired number of EOSs each of which are thermodynamically stable and follow the causal limit. Our EOSs were checked to satisfy the following : 
	\begin{itemize}
		\item The EOSs generated were checked to satisfy the causality condition imposed by the speed of sound ($0 < c_s < 1$) and also the hydrostatic stability condition, which states that the pressure should increase along with an increase in energy density.
		\item We have also taken the maximum mass limit which is obtained from the mass-radius curve by solving the Tolman-Oppenheimer-Volkoff (TOV) equations \citep{oppenheimerVolkoff1939}. The EOSs that did not have a maximum mass of at least $2 M_{\odot}$ were rejected \citep{Cromartie,Antoniadis,Fonseca}. 
		
		\item The EOSs also follow the binary tidal deformability constraints from GW170817 LIGO/Virgo. With the use of precise chirp mass measurement $\mathcal{M}_{chirp} = 1.186 M_{\odot}$ \citep{Abbot_2018,Abbot_2019} with a mass ratio $q = M_{2}/M_{1} > 0.73$ and binary tidal deformability $\Tilde{\Lambda} < 720$ with :
		\begin{equation}
			\Tilde{\Lambda} = \frac{16}{13} \left[ \dfrac{(M_{1} + 12M_{2}) M_{1}^{4} \Lambda(M_{1})}{(M_{1} + M_{2})^{5}}  + (M_{1} \xleftrightarrow{} M_{2}) \right]
		\end{equation}

	\end{itemize}

	\section{Quasi-Normal Modes of Neutron Stars}\label{sec:QNM}
	
	QNMs arise due to perturbations of stellar matter and the spacetime metric, and an angular decomposition of these perturbations into spherical harmonics will contain even and odd parity components. In this work, we are interested in QNMs (in the fully general relativistic formalism) arising from fluid perturbations that couple to gravitational waves, and thus we will restrict our focus to the dominant quadrupolar ($l=2$), even parity perturbations of the Regge-Wheeler metric \cite{thorneNonRadialPulsationGeneralRelativistic1967}:
	\begin{multline}
		\mathrm{d}s^2= -e^{2\Phi(r)}[1+r^lH_0(r)\mathcal{Y}_{lm}e^{i\omega t}]\mathrm{d}t^2\\- 2i\omega r^{l+1}H_1(r)\mathcal{Y}_{lm}e^{i\omega t}\mathrm{d}t \mathrm{d}r\\+ e^{2\Lambda(r)}[1-r^lH_0(r)\mathcal{Y}_{lm}e^{i\omega t}]\mathrm{d}r^2\\+ r^2[1-r^lK(r)\mathcal{Y}_{lm}e^{i\omega t}][\mathrm{d}\theta^2+ \sin^2\theta \ \mathrm{d}\phi^2]
	\end{multline}
	
	Here $H_0$, $H_1$, and $K$ are the perturbation functions, $\mathcal{Y}_{lm}$ are the spherical harmonics, and $\omega$ is the complex QNM frequency, the real part of which is the angular frequency of the mode, and the inverse of the positive imaginary part, the damping time. $\Phi$ and $\Lambda$ are the metric potentials corresponding to the TOV solutions of a spherically symmetric star. The perturbations of the fluid inside the star are governed by the fluid Lagrangian displacement vector, taken as 
	\begin{multline}
		\xi^i= \{r^{l-1}e^{-\Lambda}W(r), -r^{l-2}V(r)\partial_\theta,\\-r^{l-2}\sin^{-2}\theta V(r)\partial_\phi\}\mathcal{Y}_{lm}(\theta,\phi) e^{i\omega t} \label{eq:disp_vec_GR}
	\end{multline}
	
	where $W$ and $V$ are the fluid perturbation amplitudes. There exist several approaches using which one could find the QNM frequencies, such as resonance matching \cite{NonradialPulsation3Thorne1969, chandrasekhar1991}, the method of continued fractions \cite{sotaniDensityDiscontinuityNeutron2001}, WKB \cite{kokkotasWModes1992}, etc. In this work, we employ the method of direct numerical integration \cite{lindblom1983quadrupole,detweiler1985nonradial,lujunli_ChinPhyB} to find the oscillation frequencies and damping times. For completeness, we provide the equations that need to be solved and mention the numerical techniques we used to solve the same in the appendix \ref{sec:appendix1}.

	Throughout this work, the various modes are separated by the number of radial nodes \cite{coxNonradialOscillationsStars1976,rodriguezThreeApproachesClassification2023}.

	\section{Results} \label{sec:results}

\begin{figure}[!ht]
		\centering
		\includegraphics[width=\columnwidth]{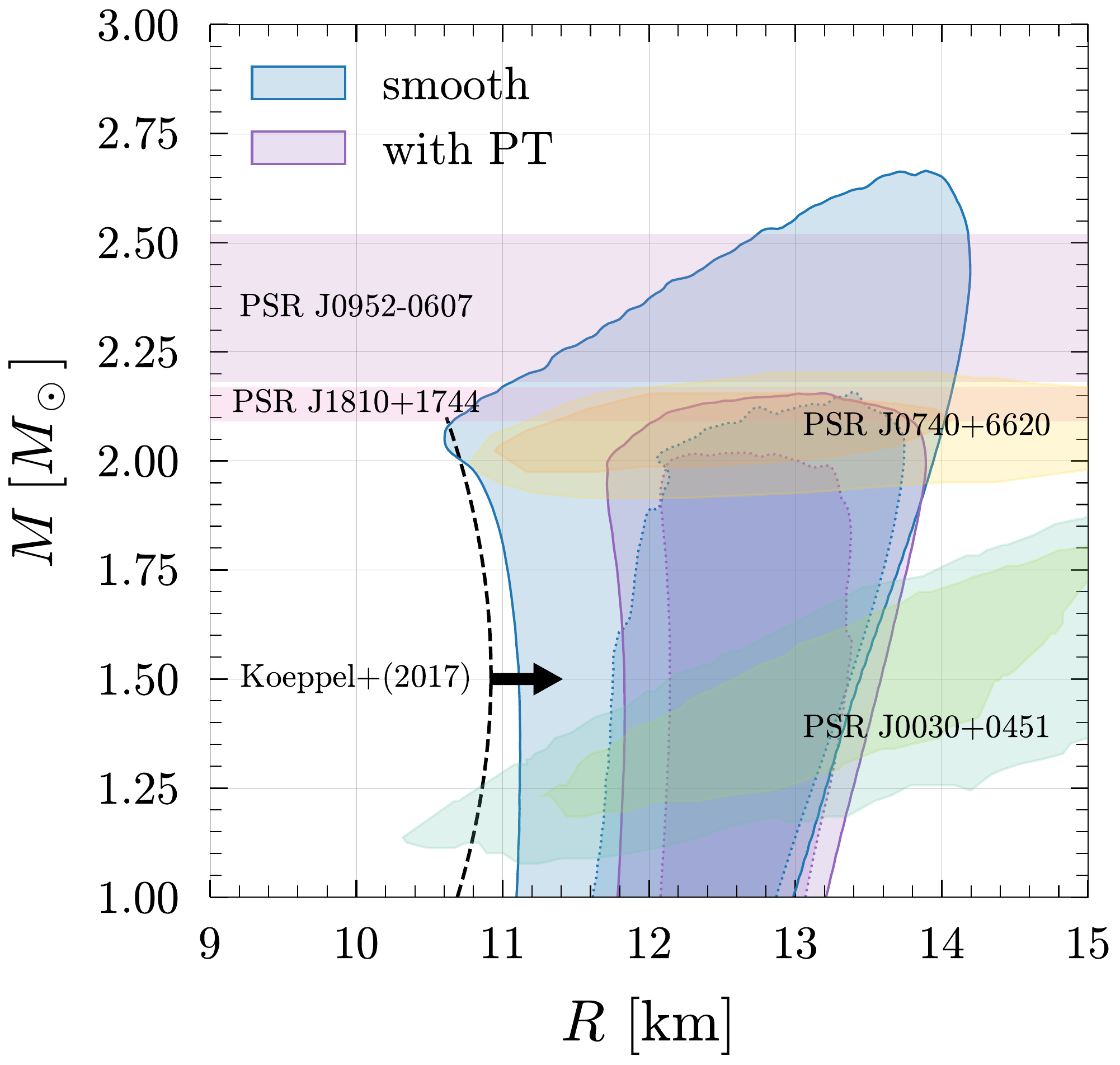}
		\caption{65\% (dotted) and 95\% (solid) contours of the mass-radius relation for the two EOS ensembles. The black dotted line indicates the Koeppel limit denoting maximum threshold mass to prompt collapse \citep{Koeppel}, and the shaded regions denote the various constraints from astrophysical observations.}
		\label{fig:MR}
	\end{figure}

	The EOSs were ensured to follow the observational constraints shown  \cref{fig:MR}, and have a maximum mass of at least 2 $M_\odot$. The 65\% and 95\% confidence contours of the M-R curve is marked from their respective probability density functions (PDFs).
	
	As clear from the M-R curves the two ensemble of EOSs marks two distinctly different contours in the M-R diagram. At low densities the EOSs with first-order PT are stiff compared to the smooth EOSs (otherwise they cannot satisfy the maximum mass constraint of $2 M_{\odot}$). Thus, at lower masses, they have comparatively higher radii \citep{gorda2023}. If the PT EOSs are to have softer EOSs at lower densities than after the PT, to satisfy the $2 M_{\odot}$ bound, they need to be uncharacteristically stiffer, thus violating the causality limit. However, for the smooth EOSs this is not the case as it can initially be softer and later on can be stiff (as they are always continuous) thereby having smaller radius at low mass values. Also, as there is no density discontinuity they can be relatively stiffer (their stiffness can vary greatly) at high densities and thus generate much higher masses. 
	
	\begin{figure}[!ht]
		\centering
		\includegraphics[width=\columnwidth]{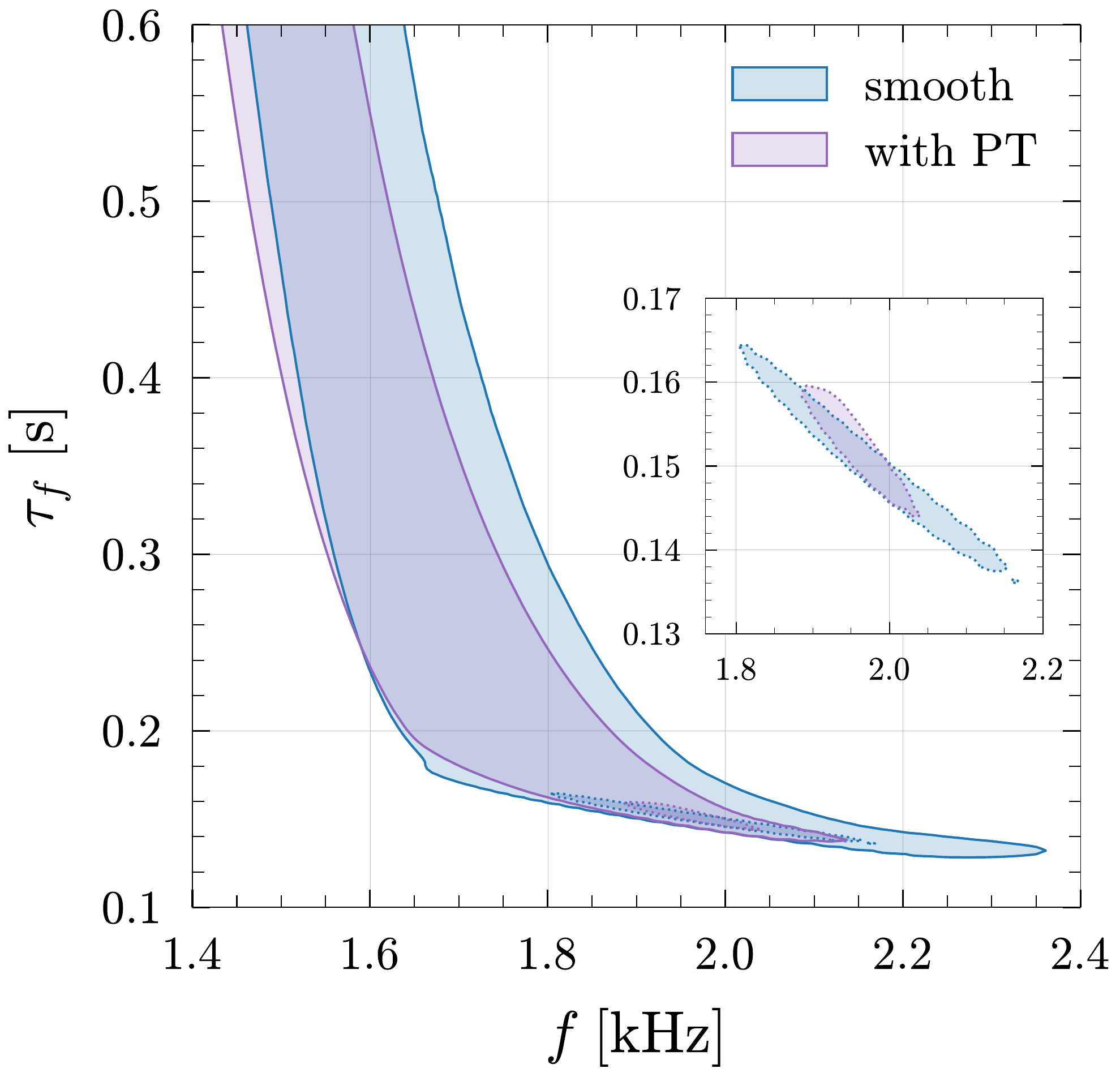}
		\caption{65\% (dotted) and 95\% (solid) contours of the damping time-frequency relation of the $f$-mode for the two EOS ensembles}
		\label{fig:FTF}
	\end{figure}
	
	\begin{figure}[!ht]
		\centering
		\includegraphics[width=\columnwidth]{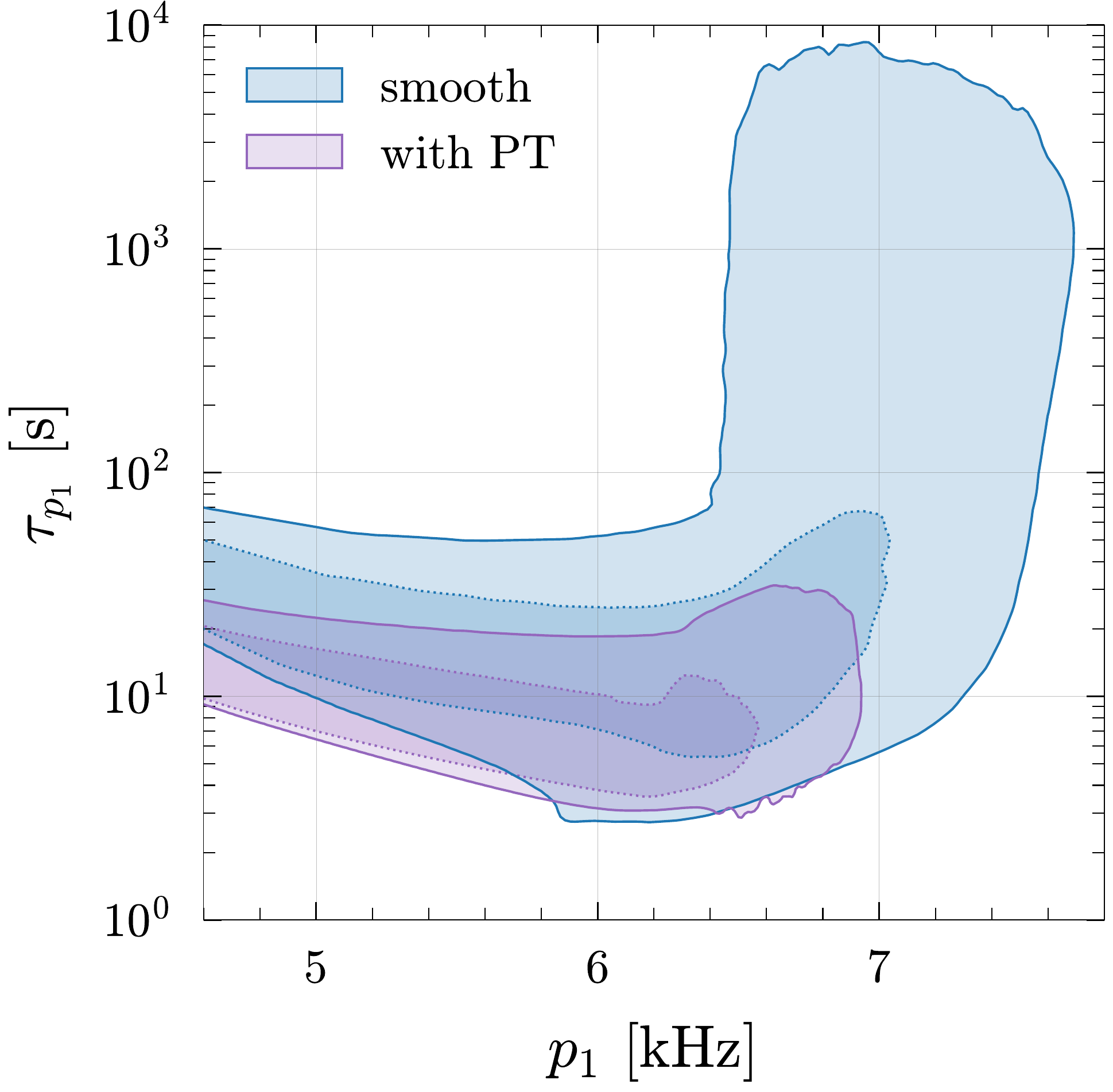}
		\caption{65\% (dotted) and 95\% (solid) contours of the damping time-frequency relation of the $p_1$-mode for the two EOS ensembles}
		\label{fig:P1TP1}
	\end{figure}
	
	The characteristic differences in the EOS revealed in the M-R curve are not limited to it. The quasi-normal modes (QNM) ($f$-modes, $p$-modes) also reveal the characteristic feature of the M-R curve of the EOSs. The $f$-mode frequencies for NSs lie at around $1.8$ kHz whereas the $p$-modes frequencies are relatively higher at around $6$ kHz. Along with the $f$-mode and $p$-mode frequency, the damping time of the frequencies is also dependent on the EOS. As already seen the M-R contours for the two distinctively different EOS are quite different, it is expected that they would also show up in the QNMs and their damping time. 
	

	In \cref{fig:FTF} the plot of damping time against frequency for $f$-mode oscillation is shown. At low frequencies and high damping times there is a difference in the two distinctively different EOS contours; however, this reduces with an increase in the frequency. There is a difference in both the $95\%$ contours and $65\%$ contours. At very high frequencies the contour for both EOS ensembles narrow down. This is due to the fact that we have set a lower mass cut off range for the stars (from observational bounds). For higher masses (beyond $2 M_{\odot}$) the number of EOSs reduce proportionally, narrowing down the contours. The incapability of EOSs with first-order PT to produce massive stars (as compared to the smooth EOSs) ends the contour for the former earlier. 
	
	
	
	The exclusion regions are more pronounced for the damping time against the frequency plot for $p_1$-modes \cref{fig:P1TP1}. At low frequencies, the overlap of both $95\%$ contours and $65\%$ contours is much less pronounced as compared to the $f$-modes. Further, the contours differ extensively at higher frequencies and higher damping times. At higher damping times only the smooth EOS contour exists. This is as due to the presence of a radial node; p-modes are more sensitive to the distribution of matter inside the star, which is in turn governed by the EOS \cite{Andersson_towards_GW_astroseismo}. Since the smooth EOSs are much less constrained by the stiffness of the EOS (the variation of stiffness at comparatively higher densities is larger compared to EOS with PT), they can produce more massive stars, and several EOS reach these large values of damping times. While, due to the constraints on the construction of EOSs with PT, very few reach such large values and these don't show up in the 95\% confidence interval.

	\subsection*{Universal relations}
	
	\begin{figure*}[!ht]
		\centering
		\includegraphics[width=\textwidth]{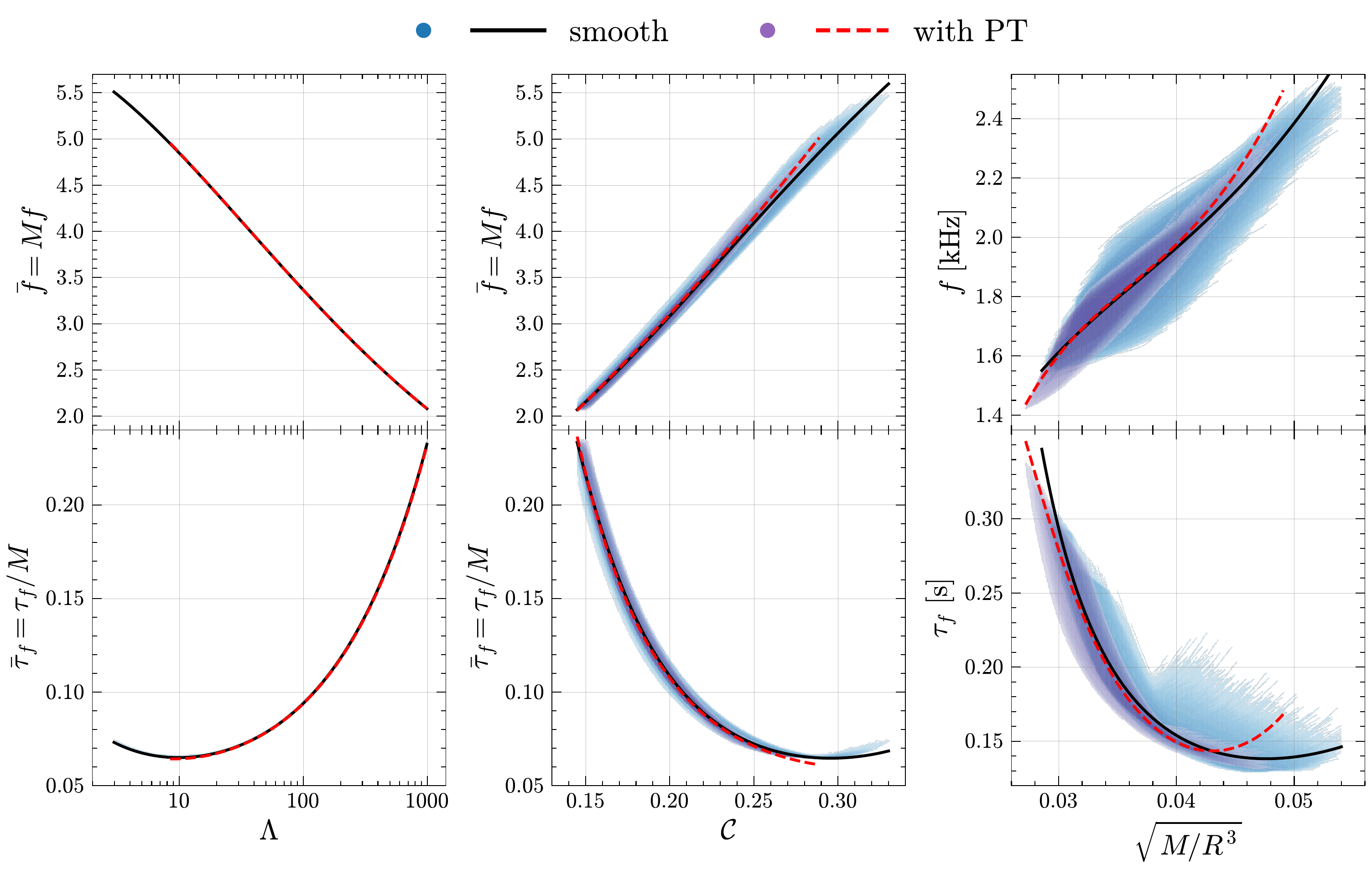}
		\caption{Relations between the dimensionless $f$-mode frequency and the dimensionless damping time with tidal deformability and compactness, and of the unscaled $f$-mode frequency and its damping time with density}
		\label{fig:UNI_LCD}
	\end{figure*}
	
	Recently, there has been extensive use of URs to study matter properties in NSs. It has led to a plethora of URs like I-love-Q, f-love, f-compactness. Usually, these are relations between dimensionless quantities from NS observables like mass, radius, love number, and compactness and are, therefore, insensitive to the details of the EOS. Because of this insensitivity, knowing one parameter is good enough to predict an unknown parameter with which it shares an UR, independent of the microphysics. However, they are quite ineffective in understanding the details of the EOS. 
	
	
	To study quasi-URs of the $f-$mode frequency and its damping time one scales these quantities appropriately. The scaling relations are: 
	
	\begin{align}
		\bar{f} &\equiv {M[M_\odot] f[\mathrm{kHz}]}\\
		\bar{\tau} &\equiv {\tau_f[\mathrm{s}]/M[M_\odot]}
	\end{align}
	
\begin{figure}[!ht]
		\centering
		\includegraphics[width=\columnwidth]{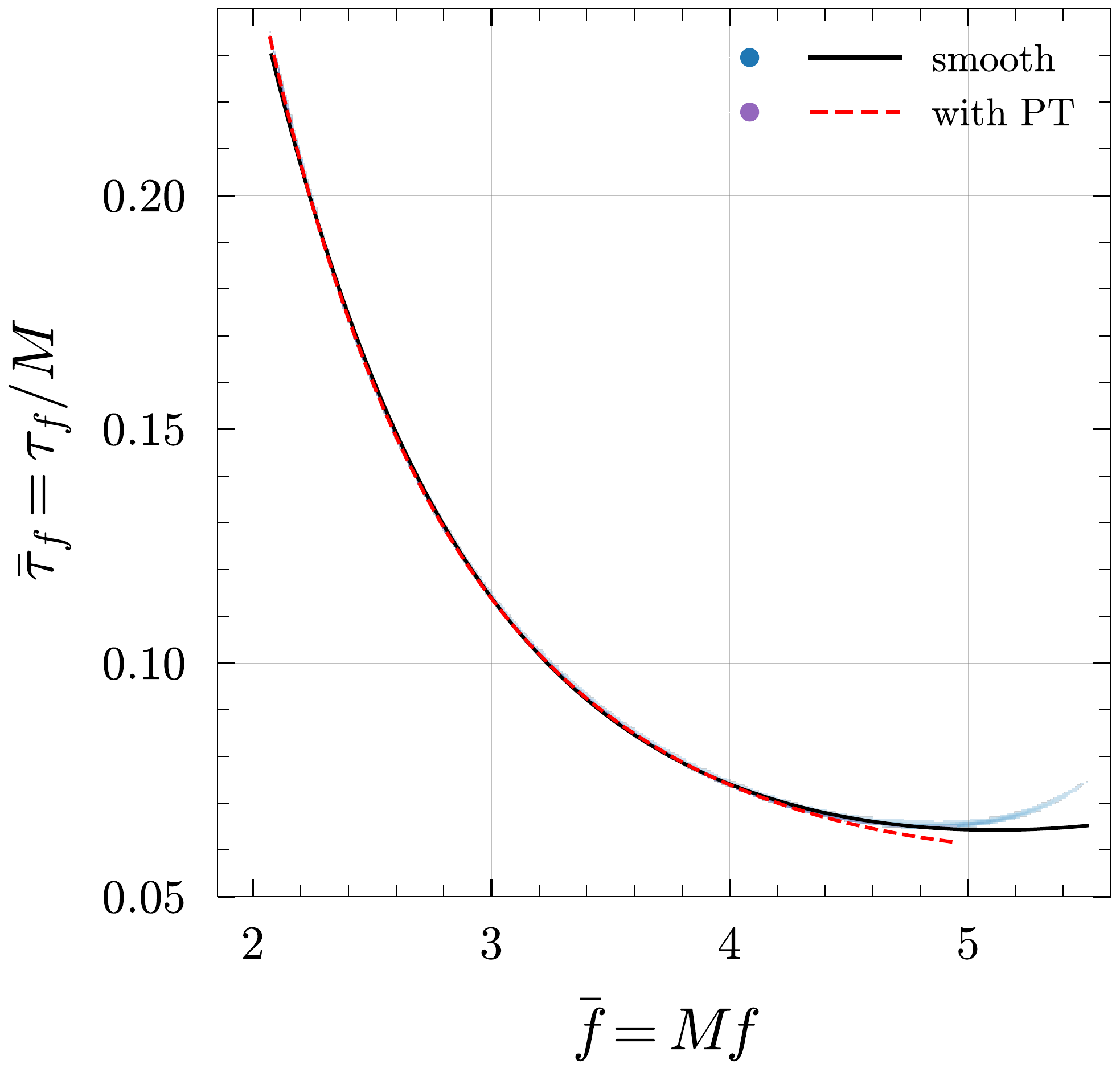}
		\caption{Relation between the dimensionless damping time with the dimensionless $f$-mode frequency}
		\label{fig:UNI_FTF}
	\end{figure}

	Universality of dimensionless f-mode frequency ($\bar{f}$) and damping time ($\bar{\tau}$)is studied with the dimensionless tidal deformability ($\Lambda$), compactness ($C$) and the average density of the star ($\sqrt{M/R^3}$) in \cref{fig:UNI_LCD}. We also study the universality between $\bar{f}$ and $\bar{\tau}$ in \cref{fig:UNI_FTF}.
	
	\begin{figure*}[!ht]
		\centering
		\includegraphics[width=\textwidth]{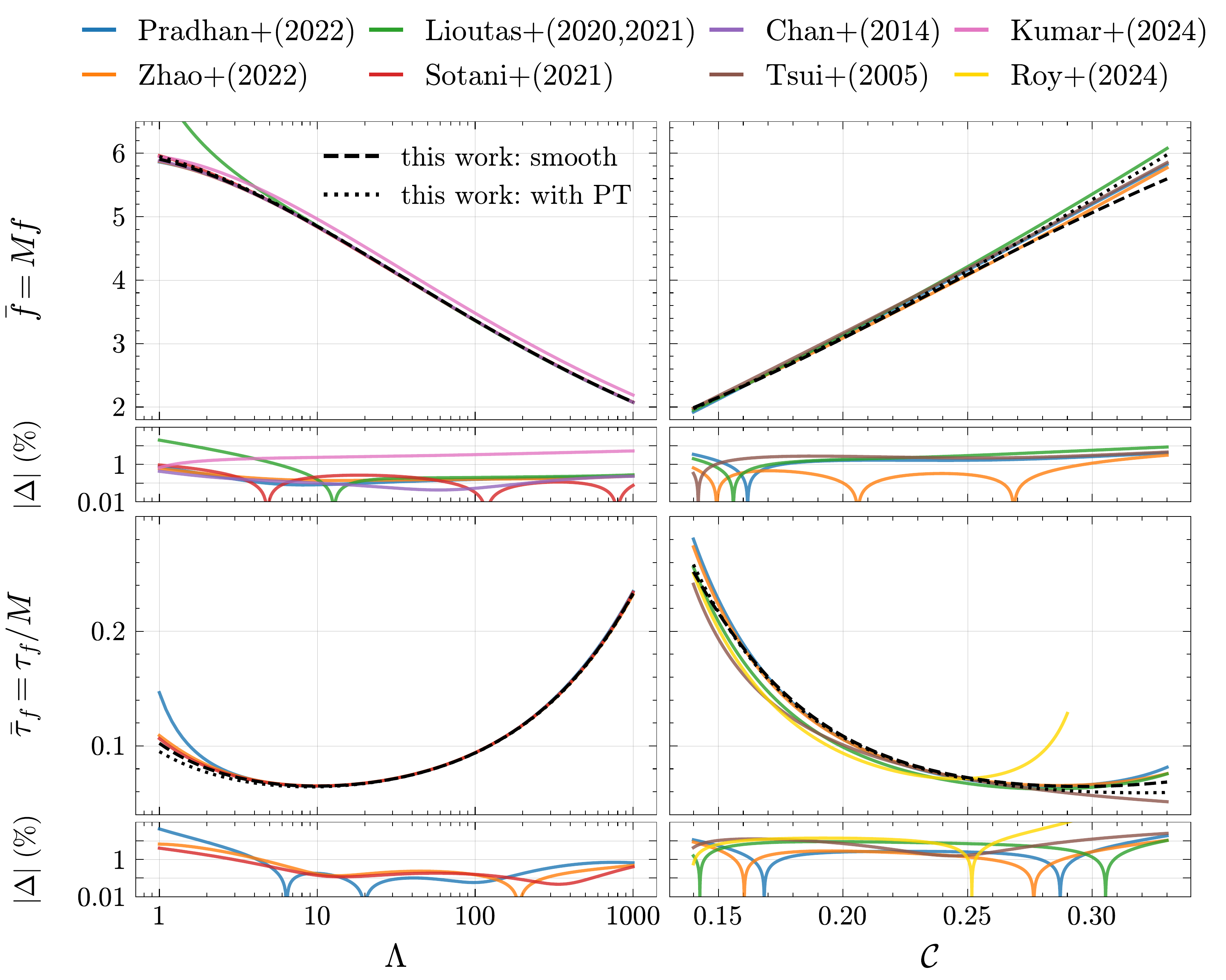}
		\caption{Comparison of universal relations with other works. Fitting functions given in \cite{Pradhan_2022} (Pradahan), \cite{zhaoUniversalRelationsNeutron2022} (Zhao), \cite{Lioutas_2020,Lioutas_2021} (Lioutas), \cite{Sotani_Kumar_2021} (Sotani), \cite{chan_2014} (Chan), \cite{LKTsui_2005} (Tsui), \cite{Kumar2023_universal_exotic} (Kumar) and \cite{roy2024analysis} (Roy) are compared with the fits in this work. The deviations from our smooth fit are shown in the smaller panels.}
		\label{fig:UNI_COMPARE}
	\end{figure*}
	
	The log scaled polynomial fitting function used is given by:
	\begin{equation}
		\log y_i = a_i + b_i \log x_i + c_i (\log x_i)^2 + d_i (\log x_i)^3
		\label{fit}
	\end{equation}
	where the coefficients are summarized in \cref{fit-tab}. 
	
	The presence of quasi-universal relations is a purely numerical phenomenon. It's seen that making specific quantities dimensionless in geometrised units removes the dependence on the EOS, and reduces the spread of the variables. Consequently, using the fitting functions, one can find the value of an unknown quantity (QNM frequency and damping time) if another quantity is known ($\Lambda$ or compactness). In the left panel, both the $x$ ($\Lambda$) and $y$ axis ($\bar{f}$ and $\bar{\tau}$) quantities are dimensionless and therefore give a very robust UR without any spread. There is no way to differentiate the two sets of EOS from this UR, as both follow the same relation.
	
	In the middle panel, again, the $x$ ($C$) and $y$-axis are dimensionless; however, in this case, there is a small spread in both the smooth and PT EOS sets. At low frequency (and high damping time), the best fit for the UR relation for the two different ensembles of EOS overlaps; however, they deviate slightly at higher frequency (low damping time).
	
	For quantities that are not dimensionless ($f$, $\tau_f$ and $\sqrt{M/R^3}$), like the rightmost panel of \cref{fig:UNI_LCD}, considerable deviations from the best-fit line exist. Also, the best fit for two different sets of EOS do not always overlap. Interestingly, universality is only prominent in the $f$-modes whereas the $p$-modes do not show universal relations. This violation of universality can be attributed to the fact that they are more sensitive to the matter distribution inside the star, making them more dependent on the EOS.  
	
	On the other hand the $f$-modes are manifestations of the average density in the star, which makes them less sensitive to the details of the EOS, and allowing their mass-scaled quantities to show minimum deviations from universality.
	
	Since the mass scaled $f$-mode frequency ($\bar{f}$) and damping time ($\bar{\tau}$) individually show universality with $\Lambda$ and $C$, it is imperative that they also show a universality with each other. This is shown in \cref{fig:UNI_FTF}.
	The mass scaled universal relations have been explored in several previous works which used nuclear EOSs. Our agnostically generated smooth EOSs are supposed to be a superset of these EOSs, and \cref{fig:UNI_COMPARE} we show the comparison between our fits and that of other previously reported UR. The amount of deviation of our best-fit from other works are shown below each plot. The EOSs generated by us are generic and are independent of any particular nuclear model, as a result they deviate from others. The coefficients used for fitting our EOSs (both smooth and PT) are shown in \cref{fit-tab}. As the fits from the PT EOSs are similar to those of smooth EOSs, hence the percentage error for them are also comparable.
	
	\begin{table*}[ht!]
		\caption{Estimated numerical coefficients for universal relation fits}
		\centering
		\begin{tabular}{@{\hspace{0.8cm}}c@{\hspace{0.8cm}}c@{\hspace{0.8cm}}c@{\hspace{0.8cm}}c@{\hspace{0.8cm}}c@{\hspace{0.8cm}}c@{\hspace{0.8cm}}c@{\hspace{0.8cm}}c}
			\hline\hline
			\noalign{\smallskip}
			Type & $y_i$ & $x_i$ &{$a_i$} &{$b_i$}
			& {$c_i$} & {$d_i$} & Error\%\\&&&&&&&Max (90\%)  \\
			\hline
			\noalign{\smallskip}
			
			Smooth & $\bar{f}  $ & $\Lambda$ & 0.7714 & -0.04235 & -0.04707 & 0.003596 & 0.23 (0.10)\\[2ex]
			with PT & $\bar{f}  $ & $\Lambda$ & 0.7744 & -0.04671 & -0.04495 & 0.00326 & 0.55 (0.30)\\[2ex]
			
			Smooth & $\bar{\tau}_f$ & $\Lambda$ & -0.9902 & -0.416 & 0.2394 & -0.02033 & 2.21 (0.30)\\[2ex]
			with PT & $\bar{\tau}_f$ & $\Lambda$ & -1.022 & -0.3705 & 0.2181 & -0.01715 & 1.24 (0.60)\\[2ex]
			
			Smooth & $\bar{f}  $ & $\mathcal{C}$ & 0.6769 & -1.783 & -4.418 & -2.121 & 6.63 (2.80)\\[2ex]
			with PT & $\bar{f}  $ & $\mathcal{C}$ & 1.412 & 1.289 & -0.1179 & -0.1122 & 5.46 (1.80)\\[2ex]
			
			Smooth & $\bar{\tau}_{f}$ & $\mathcal{C}$ & 3.832 & 23.007 & 33.169 & 14.407 & 11.41 (4.60)\\[2ex]
			with PT & $\bar{\tau}_{f}$ & $\mathcal{C}$ & 1.775 & 14.187 & 20.584 & 8.444 & 9.95 (3.30)\\[2ex]
			
			Smooth & $f$ & $\sqrt{M/R^3}$ & 20.827 & 41.24 & 27.989 & 6.435 & 12.92 (6.00)\\[2ex]
			with PT & $f$ & $\sqrt{M/R^3}$ & 51.789 & 105.394 & 72.299 & 16.636 & 8.27 (3.80)\\[2ex]
			
			Smooth & $\tau_{f}$ & $\sqrt{M/R^3}$ & 14.272 & 23.477 & 9.546 & 0.3369 & 54.42 (7.80)\\[2ex]
			with PT & $\tau_{f}$ & $\sqrt{M/R^3}$ & 150.459 & 306.473 & 205.458 & 45.524 & 21.86 (7.30)\\[2ex]
			
			Smooth & $\bar{\tau}_{f}$ & $\bar{f}  $ & -0.3968 & 0.9765 & -7.507 & 6.414 & 14.45 (0.60)\\[2ex]
			with PT & $\bar{\tau}_{f}$ & $\bar{f}  $ & -0.1201 & -0.8404 & -3.628 & 3.712 & 5.57 (0.20)\\[2ex]
			\noalign{\smallskip}
			\hline\hline
		\end{tabular}
		\label{fit-tab}
	\end{table*}

	\subsection*{Prospects of GW detection}
	\BFF{Below, we estimate the prospects of having a GW detection for the $p_1$ mode. As shown by \citep{Kokkotas, Flores_2018}, the energy released by the GW can be given as:
		\begin{align} \label{Egw}
			\dfrac{E_{GW}}{M_{\odot} c^{2}} = &3.471 \times 10^{36} \left(\dfrac{S}{N}\right)^{2} \dfrac{1+4Q^2}{Q^2} \left(\dfrac{D}{10 \text{kpc}}\right)^2 \left(\dfrac{f}{1 \text{kHz}}\right)^2 \nonumber \\ 
			&\left(\dfrac{S_n}{1\text{Hz}^{-1}}\right)
		\end{align}
		We analyze the energy required to be emitted to obtain a signal-to-noise ratio (S/N) $>$ 5. In \cref{Egw} $Q =\pi f \tau$ where $\tau$ is the damping time, $f$ is the frequency, and $S_n$ denotes the spectral noise density. From the exclusion region shown in \cref{fig:P1TP1}, we select a point with damping time ($\tau$) equal to $10$ s and frequency ($f$) equalling $7$ kHz.
		
		\begin{table*}
			\centering
			\caption{ Table showing the prospects of detection of the frequencies of $p_1$ mode with Advanced LIGO/VIRGO and Einstein Observatory (with signal to noise ratio greater than 5).}
			\label{table_GW}
			\begin{tabular}{@{\hspace{0.8cm}}c @{\hspace{0.8cm}}c @{\hspace{0.8cm}}c}
				\hline
				\hline
				& \textbf{\textit{Advanced LIGO/VIRGO}} & \textbf{\textit{Einstein Observatory}} \\ \hline  \\
				$p_1$ [kHz] & $7$ &   $7$  \\  \\ \hline \\
				$\tau_{p_1}$ [s]&  $10$ & $10$ \\ \\  \hline \\
				Spectral Noise Density ($S_{n}^{1/2}$) [Hz$^{-1}$] &  $2 \times 10^{-23}$ & $1 \times 10^{-24}$ \\ \\  \hline \\
				$E_{GW} (\times M_{\odot} c^{2})$ for D = 15 Mpc & $4$ & $9 \times 10^{-3}$ \\ \\ \hline \\
				$E_{GW} (\times M_{\odot} c^{2})$ for D = 10 kpc & $2 \times 10^{-6}$ & $4 \times 10^{-9}$ \\ \\ \hline \\
				$E_{GW} (\times M_{\odot} c^{2})$ for D = 100 kpc & $2 \times 10^{-4}$ & $4 \times 10^{-7}$ \\ \\ \hline\hline \\
			\end{tabular}
			
		\end{table*}

		In \cref{table_GW}, we show the results. The energy radiated as GW during core collapse is $\sim 10^{-5} - 10^{-6} M_{\odot} c^{2}$. For NSs lying outside our galaxy, similar to the Vela cluster, which lies at an approximate distance of 15 Mpc, the GW signal cannot be detected using either of the detectors. However, GW signals from NSs in our galaxy can be detected using Advanced LIGO/VIRGO or the third-generation Einstein Observatory. Since the estimated GW radiation at a distance of 10 kpc for both detectors lies well within a range such that we obtain a $S/N > 5$, we can say the signals from $p_1$ oscillation can be detected by both the detectors. At a distance of 100 kpc, only the Einstein observatory can detect these signals due to its higher sensitivity compared to the Advanced LIGO/VIRGO.}

	\section{Summary and Discussion} \label{sec:summary}
	
	Oscillation modes of NS has proven to be a tool of paramount importance in recent years. QNMs in NSs are excited by various mechanisms like accretion, core collapse, and tidal forces due to a close encounter; and can be probed using GWs. The $f$, and $p$ - modes having frequencies in the range of a few kHz are of utmost importance due to their possibility of detection in the present and the future detectors.
	
	
	
	This work analyses the possibility of probing the internal composition of NSs using $f$ and $p_1$ modes. The dependence of the presence or absence of a first-order PT on the $f$ and $p_1$ modes are studied. To have an unbiased check of this difference it is necessary to have an exhaustive ensemble of EOS both with and without PT. Two separate ensembles of EOSs are created in an agnostic way that satisfies the astrophysical bounds and is thermodynamically consistent. 
	
	
	Solving the TOV equations for the ensemble of EOS, one gets a broad contour in the mass-radius curve (plotting $65\%$ and $95\%$ confidence contours). The contours for smooth EOSs and for PT EOSs differ considerably. The smooth contour spreads to much larger masses and lower radii than PT contours. This indicates that the PT EOSs are stiffer than smooth EOS at low densities to fulfill the causal criterion at large densities. This feature is also manifested in their damping time and frequency. Exclusive regions exist for the smooth and with PT EOS contours and are also more prominent for $p_1$modes. This is because $p_1$ modes are more sensitive to the matter distribution inside the NS due to the presence of a radial node. It is important to note that the difference in these frequency modes is not due to the priors used for their construction. The $100\%$ contours of both $f$ and $p_1$ mode frequencies and damping time with mass (\cref{fig:QNM_100}) shows us that the smooth EOSs engulf the entire contour of the PT EOSs. The $95\%$ and $65\%$ intervals highlight that there exists an intrinsic difference that is independent of the priors used for the construction of the EOSs. This has been discussed further in \cref{appendix:100}
	
	
	URs can largely cater to understanding macroscopic variables, but fail to capture the microphysical elements of NSs. The comparison of URs for smooth and PT EOSs shows that the exclusive region has no implications on the URs. The fitting functions for both smooth and PT case with $\Lambda$ and $C$ are similar, indicating that URs cannot be used to differentiate the two types of EOSs. It is also to be noted that the nature of universality depends largely on the dimensionless parameters. Before introducing the dimensionless quantities, the quantities were not universal, implying that the dimensionless quantities can veil several interesting properties of EOSs, thus lacking effectiveness in probing the microscopic properties of NSs. \\
	\indent Although the appearance of exclusion regions is an exciting prospect with which the likelihood of an EOS (smooth or with PT) can be predicted, they lie in the range of a few kHz. Although signal-to-noise ratio is not a problem; however, present GW detectors are not very sensitive to such high frequency, and one thus has to wait for future detectors to study the likelihood of smooth and PT EOS. 
	
	\section*{Acknowledgments}
	The authors would like to thank IISER Bhopal for providing the infrastructure to carry out the research.
	SC wants to acknowledge the Prime Minister's Research Fellowship (PMRF), Ministry of Education Govt. of India, for a graduate fellowship. RM is grateful to the Science and Engineering Research Board (SERB), Govt. of India for monetary support in the form of Core Research Grant (CRG/2022/000663). K K Nath would like to acknowledge the Department of Atomic Energy (DAE), Govt. of India, for sponsoring the fellowship covered under the sub-project no. RIN4001-SPS (Basic research in Physical Sciences). The authors would like to thank C. Ecker and L. Rezzolla for providing us with the smooth EOSs  and also for their insights and comments which have helped shape this project.
	
	\section*{Data Availability}
	The data used in this work can be accessed upon reasonable request.
	
	\bibliography{references}
	
	\pagebreak
	
	\appendix
	\renewcommand{\theequation}{\Alph{section}\arabic{equation}}
	\renewcommand{\thesection}{\Alph{section}}
	\setcounter{equation}{0}

	\section{Equations governing oscillation modes of a static star}\label{sec:appendix1}
	\subsection{General relativistic formalism}
	
	Lindblom and Detweiller \cite{lindblom1983quadrupole,detweiler1985nonradial} introduced a new fluid perturbation variable $X$, to replace $V$ in \cref{eq:disp_vec_GR}. The Lagrangian pressure variations are related to this new variable by 
	\begin{equation}
		\Delta p= -r^le^{-\Phi}X\mathcal{Y}_{lm}e^{i\omega t}
	\end{equation}
	One can solve the perturbed Einstein equation, $\delta G^{\mu\nu}= 8\pi\delta T^{\mu\nu}$, to get all the relations between the perturbation functions inside the star. To avoid potential singularities in the eigenvalue problem, Lindblom and Detweiller pick the four independent variables to be $H_1$, $K$, $W$ and $X$. The differential equations governing these variables, and the algebraic relations of $H_0$ and $V$ are given as follows:
	
	\begin{widetext}
		\begin{subequations}
			\begin{align}
				H_0 &= \left\{8\pi r^3e^{-\Phi}X- \qty[(n+1)(m+4\pi r^3 p)- \omega^2 r^3 e^{-2(\Lambda+\Phi)}]H_1\right.\nonumber\\
				&\left.+ \qty[n r- \omega^2r^3e^{-2\Phi}- \frac{e^{2\Lambda}}{r}\qty(m+4\pi r^3p)\qty(3m-r+4\pi r^3 p) ]K\right\}\times\qty{3m+n r+ 4\pi r^3 p}^{-1}\\[3ex]
				V&= \qty{X + \frac{p'}{r}e^{\Phi-\Lambda}W - \frac{1}{2}(p+\epsilon)e^{\Phi}H_0}\times \qty{\omega^2(p+\epsilon)e^{-\Phi}}^{-1}\\[3ex]
				H_1' &= \frac{1}{r}\qty[l+1+\frac{2e^{2\Lambda}}{r}m+4\pi r^2(p-\epsilon)e^{2\Lambda}]H_1+ \frac{e^{2\Lambda}}{r}\qty[H_0+K-16\pi(p+\epsilon)V]\\[3ex]
				K' &= \frac{H_0}{r}+ \frac{n+1}{r}H_1- \qty[\frac{l+1}{r}-\Phi']K- \frac{8\pi}{r}(p+\epsilon)e^{\Lambda}W\\[3ex]
				W' &= -\frac{1}{r}(l+1)W+ re^{\Lambda}\qty[\frac{e^{-\Phi}}{(p+\epsilon)}\dv {\epsilon}{p}X- \frac{2}{r^2}(n+1)V+\frac{1}{2}H_0+K]\\[3ex]
				X' &= -\frac{1}{r}lX+ (p+\epsilon)e^{\Phi}\left\{\frac{1}{2}\qty[\frac{1}{r}-\Phi']H_0+ \frac{1}{2}\qty[r\omega^2 e^{-2\Phi}+ \frac{n+1}{r}]H_1\right.\nonumber\\
				&- \frac{1}{r}\qty[4\pi(p+\epsilon)e^{\Lambda}+ \omega^2e^{\Lambda-2\Phi}-r^2\qty(\frac{e^{-\Lambda}}{r^2}\Phi')']W
				\left. + \frac{1}{2}\qty[3\Phi'-\frac{1}{r}]K- \frac{2}{r^2}(n+1)\Phi'V \right\}
			\end{align} \label{eq:fluid_eq_GR}   
		\end{subequations}
	\end{widetext}
	
	where $n=(l-1)(l+2)/2$. The system of differential and algebraic equations, \cref{eq:fluid_eq_GR} completely describes the perturbations inside the star. It's clear that the system of differential equations is singular at $r=0$, and the system blows up rapidly close to the stellar centre. To circumvent this problem, if $Y= \{H_1, K, W, X\}$, then near the center, $Y(r)$ is approximated as $Y(r)= Y(0)+ \frac{1}{2} Y''(0)r^2+ \mathcal{O}(r^4)$. The forms of $Y(0)$ and $Y''(0)$ are given in eq. (15) of \cite{lujunli_ChinPhyB}. These form the initial boundary conditions of the problem. The surface boundary condition is simply that at the surface of the star, the pressure perturbations, and thus $X$ must be $0$. To solve \cref{eq:fluid_eq_GR}, we follow the method outlined in \cite{lindblom1983quadrupole}. We start off with 3 linearly independent solutions at the surface, and 2 linearly independent solutions at the centre and integrate them backwards and forwards to some point inside the star where they are matched. A linear combination of these solutions, with the coefficients obtained after matching, gives the true values of $H_1$ and $K$ at the surface of the star. These variables; and $H_0$ are the only variables defined outside the star, where the perturbation equations reduce to the Zerilli equation \cite{lujunli_ChinPhyB,zhaoUniversalRelationsNeutron2022}:
	\begin{equation}
		\dv[2]{Z}{{r^*}}+ \qty[\omega^2- \mathcal{V}(r^*)]Z= 0 \label{eq:Zerilli_eq}
	\end{equation}
	where $\mathcal{V}(r^*)$ is the Zerilli potential, 
	
	\begin{equation}
		\mathcal{V}(r)= \qty(1-2b)\frac{2n^2(n+1)+ 6n^2b+18nb^2+ 18b^3}{r^2(n+3b)^2}
	\end{equation}
	
	$r^*$ is the tortoise coordinate, $r^*= r+ 2M\ln\qty(r/2M-1)$, $n= (l-1)(l+2)/2$ and $b=M/r$, with $M$ being the total mass of the star. 
	
	In case of a first-order with PT in the star, we impose additional junction conditions which ensure the continuity of $H_1$, $K$, $W$ and $X$ across the point of discontinuity of energy density \cite{sotaniDensityDiscontinuityNeutron2001}.
	
	The perturbed metric outside the star describes a combination of outgoing and incoming gravitational waves, which is the general solution to the Zerilli equation. We are interested in the case of purely outgoing waves, representing the QNMs of the star. At the surface of the star, where $r=R$, the fluid variables can be converted to the Zerilli ones using \cite{zhaoUniversalRelationsNeutron2022}:
	
	\begin{subequations}
		\begin{align}
			Z(R)&= \frac{R\qty[k(R)K(R)-H_1(R)]}{k(R)g(R)-h(R)}\\[3ex]
			\dv{Z(r^*)}{r^*} &= \frac{-h(R)K(R)+g(R)H_1(R)}{k(R)g(R)-h(R)}
		\end{align}\label{zerilli_far}
	\end{subequations}
	Here
	
	\begin{subequations}
		\begin{align}
			g(r)&= \frac{n(n+1)+3nb+6b^2}{n+3b}\\
			h(r)&= \frac{n- 3nb- 3b^2}{(1-2b)(n+3b)}\\
			k(r)&= 1/(1-2b) 
		\end{align}
	\end{subequations}
	
	After continuing the integration of the Zerilli equation \cref{eq:Zerilli_eq} to sufficiently far away from the star ($\sim 50\omega^{-1}$), the solution can be approximated as a linear combination of incoming and outgoing waves as $Z(r^*)= A_-(\omega)Z_-(r^*)+ A_+(\omega)Z_+(r^*)$ where $Z_-$ represents the outgoing wave, $Z_+$ the incoming wave and $A_-$ and $A_+$ their amplitudes. At a large enough radius, 
	
	\begin{subequations}
		\begin{align}
			Z_-&= e^{-i\omega r^*} \qty[\beta_0+\frac{\beta_1}{r}+\frac{\beta_2}{r^2}+ \mathcal{O}(r^3)]\\[3ex]
			\dv{Z_-}{r^*}&= -i\omega e^{-i\omega r^*}\qty[\beta_0+ \frac{\beta_1}{r}+ \frac{\beta_2- i\beta_1(1-2M/r)/\omega}{r^2}]
		\end{align}
	\end{subequations}
	Here $Z_+$ is the complex conjugate of $Z_-$ (and hence $A_+$ the complex conjugate of $A_-$) and, \cite{zhaoUniversalRelationsNeutron2022}
	\begin{subequations}
		\begin{align}
			\beta_1&= {-i(n+1)\beta_0}/{\omega}\\
			\beta_2&= {\qty[-n(n+1)+ iM\omega(3/2+ 3/n)]\beta_0}/{2\omega^2}
		\end{align}
	\end{subequations}
	$\beta_0$ can be any complex number that represents an overall phase. By matching the solution of $Z(r_*)$ and ${\mathrm{d}Z(r^*)}/{\mathrm{d}r^*}$ obtained from \cref{zerilli_far} with the above equation, we can find the amplitude $A_+$ with a simple matrix inversion \cite{zhaoUniversalRelationsNeutron2022}. The frequency of the QNM corresponds to that $\omega$ which minimises $A_+=0$. 
	
	To find the QNM frequency and its damping time we first find $A_+$, which in general will be a complex number, for several real values of $\omega$ close to the original guess. We then perform a complex polynomial fitting to approximate a parabola passing through the $A_+$ points corresponding to the $\omega$ values. The root of this parabola which has a positive imaginary part is the required complex $\omega$ of our QNM. We then take the real part of this $\omega$ and repeat the entire procedure several more times till the desired tolerance is reached. The real part of this final $\omega$ is the frequency of the QNM. The inverse of the imaginary part is the corresponding damping time. 
	
	For the initial guess, we use Brent's method to find all minimas of $A_+(\omega)$. By checking the number of radial nodes of the fluid variables, as well as by comparing the values of the minimas with each other, it can be found which minima corresponds to which fluid oscillation mode ($f-$ or $p_1-$). These minimas are then taken to be the initial guesses for the corrsponding mode.
	\subsection{Notes on computation}
	\begin{figure*}
		\centering
		\includegraphics[width=\textwidth]{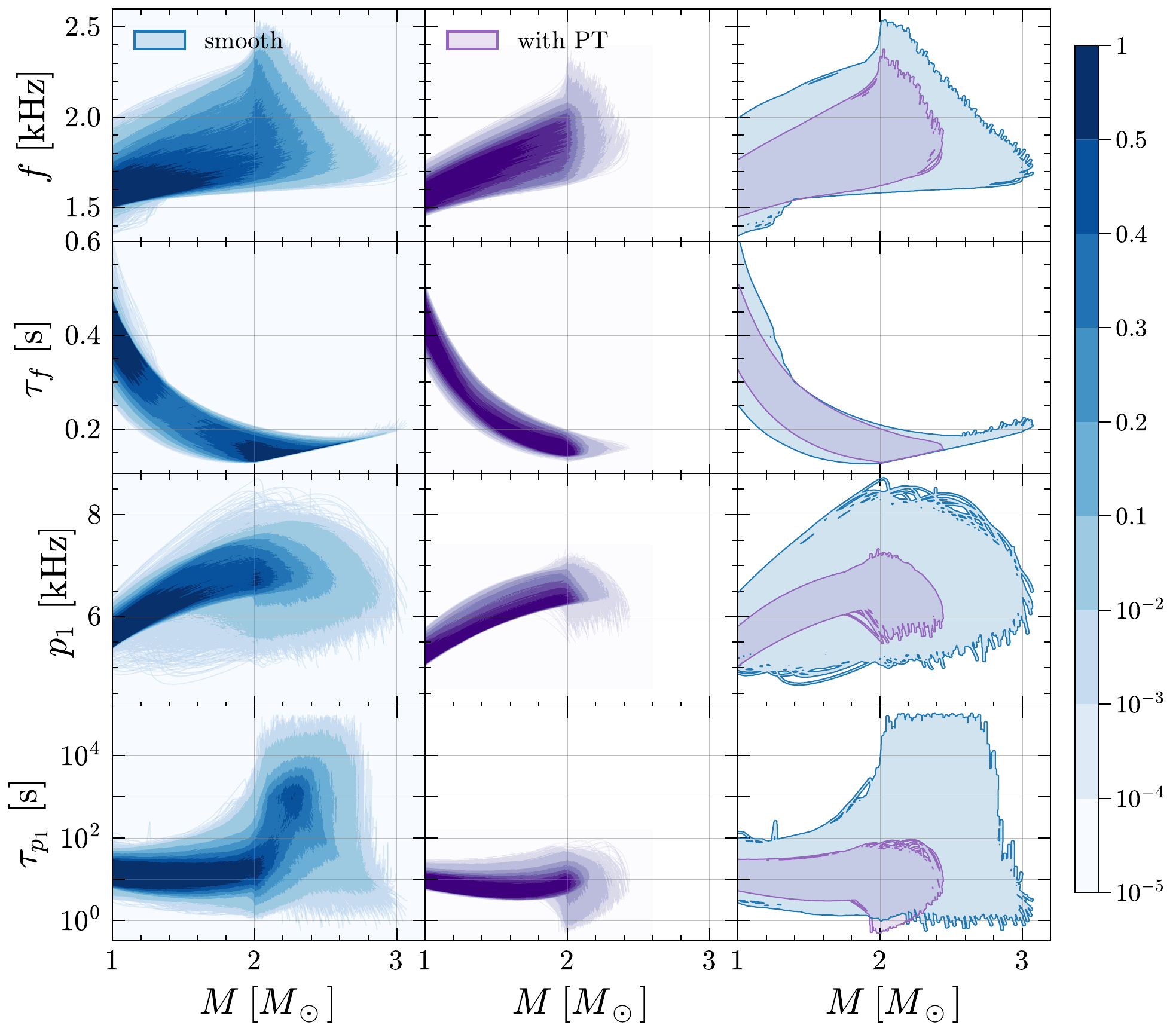}
		\caption{PDFs and 100\% contours of frequency and damping time vs mass for the $f$ and $p_1$-modes for the two ensembles.}
		\label{fig:QNM_100}
	\end{figure*}
	
	The numerical integration of all relevant ODEs was performed using an ODE solver, LSODA, which automatically adjusts the step size and switches between non-stiff (Adam's) and stiff (BDF) methods. We further used a thread-safe version of this algorithm \cite{ODEPACK_LSODA} so that we can run our code in parallel on multiple threads, significantly reducing the computation time required for the large number of EOSs in this work. We specified the relative tolerance to be $10^{-8}$ and the absolute tolerance to be $10^{-12}$ throughout the code; and obtained satisfactory results. The validity of our code was checked thoroughly by comparing our results with those in several previous works \cite{kunjipurayilImpactEquationState2022,sotaniDensityDiscontinuityNeutron2001}. Moreover, we also calculated the oscillation modes and damping times using Thorne \cite{NonradialPulsation3Thorne1969} Ferrari's \cite{chandrasekhar1991} Breit-Wigner resonance fitting approach and the frequency values matched those obtained by the Lindblom \cite{lindblom1983quadrupole,detweiler1985nonradial} approach.
	
	\section{Sampling technique} \label{appendix:100}

	The ensemble of EOSs generated in this work uses two different priors. It is essential to understand whether the distinct exclusion region for the $65\%$ and $95\%$ contours is due to our sampling technique. The contours are obtained from the corresponding probability density functions (PDFs). To find these, we divide the $x-y$ plane into a fixed-resolution grid and count the number of curves passing through each grid cell. To smoothen the distribution, we apply a 1$\sigma$ gaussian filter. Finally, we normalize the distribution by dividing all cell counts by the maximum count. The resulting PDFs and the 100\% contours for the frequency and damping time with mass are shown in \cref{fig:QNM_100}. In \cref{fig:QNM_100} (right panels) we show the entire contour corresponding to the damping time vs masses for both $f$ and $p_1$ modes. The figures show that the PT EOSs all lie within the contour of the smooth EOSs for both cases. Any existing bias that might arise due to the different sampling techniques of our EOSs would have resulted in two separate contours consisting of an exclusion region even for the 100\% 


\end{document}